\pdfoutput=1%
\documentclass[10pt,letterpaper]{article}
\usepackage{opex3}
\usepackage{amsmath,amsfonts}
\usepackage{cite}
\usepackage{etex}
\usepackage[T1]{fontenc}
\usepackage{graphicx}
\usepackage[utf8]{inputenc}
\usepackage{microtype}
\usepackage{textcomp}
\usepackage{xspace}
\usepackage{layouts}
\usepackage{soul}
\usepackage{color}
\usepackage[todos]{CMImacros}

\newcommand{\LuO}{\ensuremath{\text{Yb:Lu}_2\text{O}_3}\xspace}%
\newcommand{\YAG}{\ensuremath{\text{Yb:YAG}}\xspace}%

\newlength{\figwidth}
\begin{document}
\title{High-intracavity-power thin-disk laser for the alignment of molecules}%
\author{\small Bastian Deppe,$^{1,2,3,4}$ Günter Huber,$^{1,2,3}$ Christian Kränkel,$^{1,2,3,5}$\\
   and Jochen Küpper$^{1,2,4,6}$}%
\address{%
   $^1$The Hamburg Center for Ultrafast Imaging, University of Hamburg, Luruper Chaussee 149, 22761
   Hamburg, Germany \\
   $^2$Department of Physics, University of Hamburg, Luruper Chaussee 149, 22761 Hamburg, Germany \\
   $^3$Institut für Laser-Physik, University of Hamburg, Luruper Chaussee 149, 22761 Hamburg, Germany \\
   $^4$Center for Free-Electron Laser Science, DESY, Notkestraße 85, 22607 Hamburg, Germany \\
   $^5$ckraenkel@physnet.uni-hamburg.de \\
   $^6$jochen.kuepper@cfel.de \\
}%
\begin{abstract}
   We propose a novel approach for strong alignment of gas-phase molecules for experiments at
   arbitrary repetition rates. A high-intracavity-power continuous-wave laser will provide the
   necessary ac electric field of $\!10^{10}$--$10^{11}\;\textup{W/cm}^2$. We demonstrate thin-disk
   lasers based on \YAG and \LuO in a linear high-finesse resonator providing intracavity power
   levels in excess of 100~kW at pump power levels on the order of 50~W. The multi-longitudinal-mode
   operation of this laser avoids spatial-hole burning even in a linear standing-wave resonator. The
   system will be scaled up as in-vacuum system to allow for the generation of fields of
   $10^{11}~\text{W}/\text{cm}^2$. This system will be directly applicable for experiments at modern
   X-ray light sources, such as synchrotrons or free-electron lasers, which operate at various very
   high repetition rates. This would allow to record molecular movies through temporally resolved
   diffractive imaging of fixed-in-space molecules, as well as the spectroscopic investigation of
   combined X-ray--NIR strong-field effects of atomic and molecular systems.
\end{abstract}

\bibliographystyle{osajnl}%

\section{Introduction}
\label{sec:introduction}
Many experiments aiming at recording so-called ``molecular movies'' -- the atomic-resolution imaging
of the intrinsic structural dynamics of molecules -- rely on molecules fixed-in-space, \ie, aligned
or oriented samples of molecules~\cite{definition-alignment+orientation, Stapelfeldt:RMP75:543}. One
of the most promising approaches is coherent X-ray diffractive imaging, for instance, using
short-wavelength free-electron lasers or synchrotrons~\cite{Neutze:Nature406:752,
   Filsinger:PCCP13:2076, Barty:ARPC64:415, Kuepper:PRL112:083002}. These modern light sources,
especially upcoming free-electron lasers with very large photon fluxes such as the European
XFEL~\cite{Altarelli:XFEL-TDR:2006} or LCLS II~\cite{BESAC:XRAY:2013}, but also
synchrotrons~\cite{Franz:SynchRadNews19:25} as well as table-top laser systems based on
optical-parametric chirped-pulse amplification (OPCPA)~\cite{Vaupel:OptEng53:051507}, operate at
very high repetition rates with ten-thousands to millions of pulses per second, sometimes in burst
modes, which have to be matched by the high-intensity optical control lasers.

The alignment of gas-phase ensembles of molecules exploits the interaction between the anisotropic
polarizability of the molecule and non-resonant linearly or elliptically, polarized electric
fields~\cite{Friedrich:PRL74:4623, Stapelfeldt:RMP75:543}. The electric field strengths necessary
for strong (quasi) adiabatic alignment are on the order of
$10^{10}\text{--}10^{12}~\text{W/cm}^{2}$, even when exploiting very cold
samples~\cite{Holmegaard:PRL102:023001, Trippel:MP111:1738, Chang:IRPC34:557}, and must be applied
over durations longer than, or at least comparable to, the rotation periods of the
molecules~\cite{Trippel:PRA89:051401R}. These range from tens of picoseconds for small molecules to
nanoseconds or even microseconds for larger molecules. Adiabatic mixed-field orientation requires
the addition of a moderate dc electric field and even longer laser
pulses~\cite{Holmegaard:PRL102:023001, Nielsen:PRL108:193001, Trippel:PRL114:103003}.
Three-dimensional control requires elliptically polarized fields with fully controllable
ellipticity~\cite{Larsen:PRL85:2470, Nevo:PCCP11:9912}. Traditionally, injection-seeded Nd:YAG
lasers operating at a repetition rate of some 10~Hz were used to provide the necessary fields.
Recently, we have demonstrated the use of chirped pulses from amplified Ti:Sapphire laser systems at
1~kHz~\cite{Trippel:MP111:1738, Trippel:PRA89:051401R} and the implementation of this
amplified-chirped-pulse technique at the Linac Coherent Light Source (LCLS) at
120~Hz~\cite{Kierspel:JPB:48:204002}. Continuous-wave (CW) lasers would allow for the control of
molecules at arbitrary repetition rates. Furthermore, they would enable long interaction times. For
instance, molecules or particles traveling through a 50~\um laser beam with a velocity of
100~m/s~\cite{Benner:JAeroSci39:917, Kirian:SD2:041717, Hutzler:CR112:4803} would experience an
effective pulse duration of 0.5~\us. However, the use of CW lasers to provide the envisioned field
strengths with focal beam waists in excess of $\omega_0=10~\um$ would require optical power levels
of 30--300~kW, respectively, with a beam quality that allows for such tight focusing with a Rayleigh
length comparable to the molecular beam diameter on the order of 1~mm.

100-kW-class CW lasers have been realized as fiber lasers~\cite{Shcherbakov:ASSL2013:ATh4A.2}, as
CO$_2$ lasers~\cite{Aksakal:NIMPRS600:155} and as chemical deuterium fluoride
lasers~\cite{Mustafiz:PSaT22:1187}. Coherent beam combining ~\cite{Fan:IEEEQE11:567} is also a
viable approach to achieve laser output at this power level~\cite{Mourou:NatPhoton7:258}. The
thin-disk laser (TDL) geometry has been shown to be suited for very high CW output powers in
combinition with Yb$^{3+}$-doped gain materials~\cite{Giesen:APB58:365}. Outputs exceeding 27~kW
were demonstrated~\cite{Giesen:CLEO2013} and 100-kW-systems are anticipated~\cite{Giesen:CLEO2013,
Gottwald:SPIEProc8898}. However, so far these high output powers of thin-disk lasers and fiber
lasers are only available at beam qualities~\cite{Gottwald:SPIEProc8898,
Shcherbakov:ASSL2013:ATh4A.2} that do not allow for the tight focusing necessary to achieve the
envisioned focal intensities. In contrast, carbon dioxide and chemical lasers have good beam
qualities at even higher output power, but their demanding space requirements and the possibly toxic
gain materials are a significant drawback for using the setup as a mobile user facility at modern
X-ray light sources. Moreover, operating and propagating any laser at such high output power levels
imposes serious safety risks.

Here, we propose to provide the necessary field strength in an intracavity focus of a CW TDL
resonator. A resonator with low losses, low output coupler transmission, and high intracavity power
has a low stored excitation energy in the crystal, a high photon energy storage in the resonator and
comparably low output power. The required pump power levels are reasonably low due to the strong
enhancement in the active cavity and allow for a cost efficient system, without an exceptional
laser-safety risk. In combination with its small spatial footprint, such a system is an ideal
candidate for flexible and safe employment at modern light sources. A long multiply-folded resonator
allows for TEM$_{00}$ operation and enables to focus the beam with the necessary Rayleigh length of
1-mm.

\section[Design criteria for an high-intracavity-power Yb(3+)-based continuous-wave laser]{Design
   criteria for an high-intracavity-power Yb$^{3+}$-based continuous-wave laser}
\label{sec:design-criteria}
\begin{figure}[b]
   \centering%
   \includegraphics[width=\textwidth]{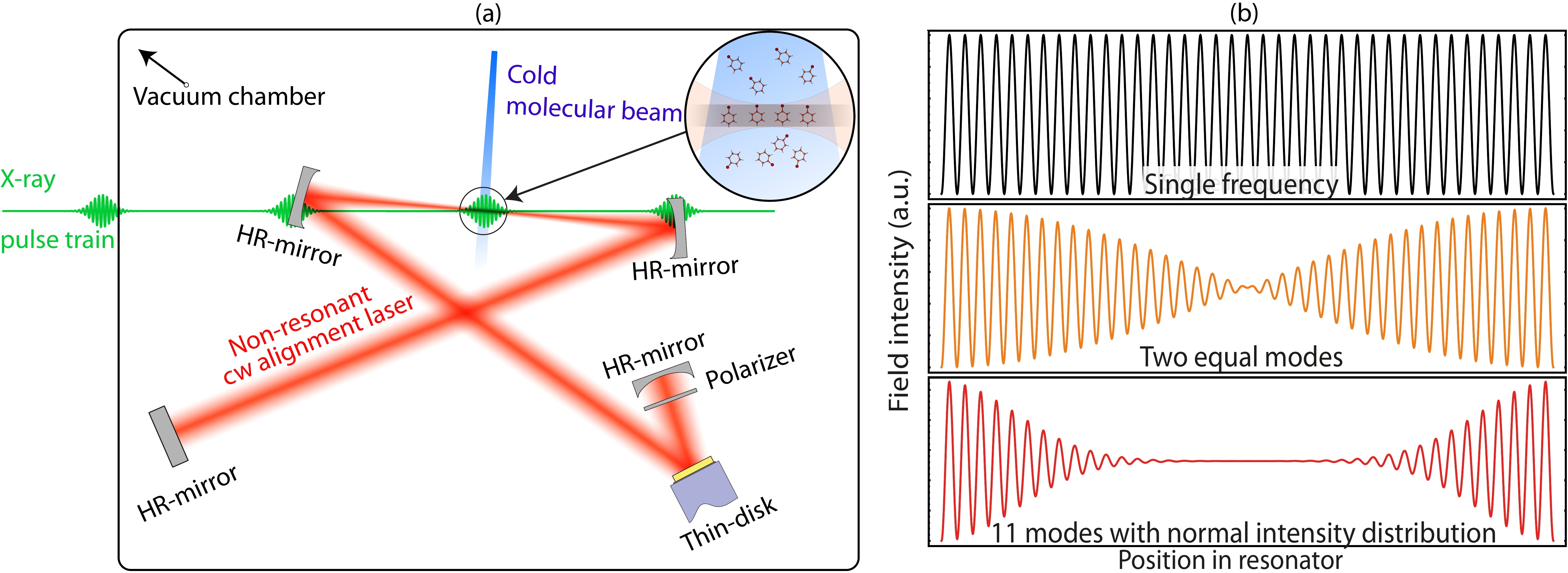}%
   \caption{(a): Concept for CW alignment of molecules: a high-finesse resonator in linearly
      polarized TEM$_{00}$ operation, focused inside of a vacuum chamber leads to the required focal
      intensities for alignment of molecules. (b): Mode averaging in the resonator yields in smooth
      intensity distribution and avoids spatial hole burning.}
   \label{fig:resonator-scheme}
\end{figure}
The envisioned aligned-molecule-imaging experiments impose a number of design criteria on the TDL.
These arise partly from the actual molecular physics and partly from the need for integration of the
alignment laser into the complex and constraining machinery of the experimental setup, schematically
shown in Fig.~\ref{fig:resonator-scheme}. The ultra-high-vacuum needs impose further restrictions
on the design of the setup.

In order to ensure the alignment of all probed molecules, the electric field intensity of
$\sim\!10^{11}~\text{W/cm}^2$ utilized to control the molecules, has to be spatially and temporally
smooth and nearly constant over the diffraction volume. The latter is defined by the overlap of the
few-mm-diameter molecular beam and the 10-\um-diameter X-ray-beam. Such a homogeneous field can be
achieved inside the TDL resonator through multi-longitudinal-mode operation, as depicted in the
inset of Fig.~\ref{fig:resonator-scheme}. The different modes average out the field distribution
around the interaction point. For a few ten modes a longitudinally practically homogeneous field is
achieved in the resonator. Under these conditions, spatial hole burning could only occur close to
the end mirrors. Multi-longitudinal mode operation does not impose a serious challenge for the
bandwidth of the laser material or the resonator design. At a typical lengths of a TEM$_{00}$ TDL of
1~m, hundreds of longitudinal modes fit into 1~pm of laser emission bandwidth in the 1~\um
wavelength range. Yb$^{3+}$-doped lasers typically exhibit emission bands with much broader
bandwidths in the nanometer range.

The laser beam needs to be focused to create the necessary field strength. At the same time, the TDL
beam needs to be larger than the X-ray beams, which are typically kept on the order of 10~\um to
avoid radiation damage~\cite{Kuepper:PRL112:083002, Stern:FD171:393}, and its Rayleigh length needs
to be long enough to provide the necessary field strength over the width of the molecular beam of
typically a few mm. This results in an envisioned focal waist of the TDL of approximately
20--40~\um. Such small focal spot sizes require sophisticated astigmatism compensation, which has to
be considered in the resonator design. Furthermore, a stable linear or elliptical polarization state
is required.

The cold molecular beam has to be delivered to the intracavity focus and the degree of alignment
needs to be monitored, for instance, through velocity map ion imaging
(VMI)~\cite{Eppink:RSI68:3477}. At the current state, the dimensions of the delivery mechanics and
the VMI device require a clear space of tens of cm$^3$ around the focus. Moreover, the X-ray beam
must pass the focus nearly collinear with the TDL intracavity mode to achieve a good overlap with
the volume of strongly aligned molecules.

Finally, the laser wavelength must be off-resonant with respect to the molecular sample to avoid
excitation and radiation damage from the control beam. As the electronic transitions of molecules
are typically in the ultraviolet or visible spectral range and vibrational excitations in the
mid-infrared range, this requirement is usually fulfilled by Yb$^{3+}$-based lasers with emission
wavelengths around 1~\um.

\section{Experimental setup}
\label{sec:setup}
In a first proof-of-principle experiment a TDL was set up in a short longitudinal resonator. This
allowed to characterize different gain materials at low output coupler transmissions with respect to
their losses and their efficiencies. The experiments were carried out utilizing a
Yb(7~\%):Y$_{3}$Al$_{5}$O$_{12}$ (Yb:YAG) disk (Dausinger+Giesen) with a thickness of 0.22~mm and a
Yb(3~\%):Lu$_{2}$O$_{3}$ disk with a thickness of 0.25~mm \cite{Peters:JCrystGrowth310:1934}. Due to
the different cation densities \cite{Gmelin:1974ue} in both host materials, the different values for
the Yb$^{3+}$-doping correspond to a similar Yb$^{3+}$-density of \textasciitilde
10$^{21}$~cm$^{-3}$. Both disks were soldered onto copper-tungsten-alloy heat-sinks (20/80 for \YAG
and 10/90 for \LuO), which were cooled at a water temperature of \celsius{6} during the experiments.
The disks were pumped by 600~\um-core multimode fiber coupled InGaAs laser diodes, which were imaged
onto a 1.2~mm diameter pump spot on the disk. For the \YAG disk a pump wavelength of 940~nm
corresponding to a broad \YAG absorption band in this wavelength range was chosen. The Jenoptik
JOLD-75-CPXF-2P laser diode utilized for this purpose had a maximum output power of 75~W and up to
56~W were used in the experiments. In contrast, \LuO provides a much stronger absorption at the
zero-phonon line around 976~nm\cite{Peters:OE15:7075}. The corresponding JOLD-50-CPXF-2P pump laser
diode provided up to 50~W of output power which was fully utilized. The temperature of the pump
diodes was adjusted to fine-tune the emission wavelength for an optimum absorption of the pump
power. After the 24 pump-light passes in our TDL module, more than 99~\% of the pump power was
absorbed. A simple plane-concave resonator with a length of 60~mm and different output coupling
mirrors with radii of curvature (ROC) of 100~mm (ROC of the disks $\approx$ 2~m) was set up for
efficient multimode laser operation. The output-coupling mirrors had transmissions between
$9.5\times10^{-5}$ and $4\times10^{-3}$ for all wavelengths between 1~\um and 1.1~\um. The output
coupler transmissions were initially measured with a spectrometer (Varian Cary 5000) and
cross-checked by measuring the transmission of the output of a \LuO TDL operating at wavelength of
1080~nm through these mirrors.
The laser output power was measured with a Coherent LM-100 power meter. To avoid damage of the disks
we limited the maximum pump power to the value which resulted in a disk surface temperature of
$\approx \celsius{120}$. For this purpose we monitored the surface temperature of the laser disks
with a SC645 thermographic camera (FLIR Systems). The measured values were corrected by the
temperature dependent emissivity of the respective gain material. The emission spectra of the lasers
were measured with a spectrometer (Ocean Optics HR2000), suitable for the wavelength range between
950~nm and 1100~nm.

\section{Experimental results}
\label{sec:results}
\begin{figure}[b]
   \centering \includegraphics[width=\textwidth]{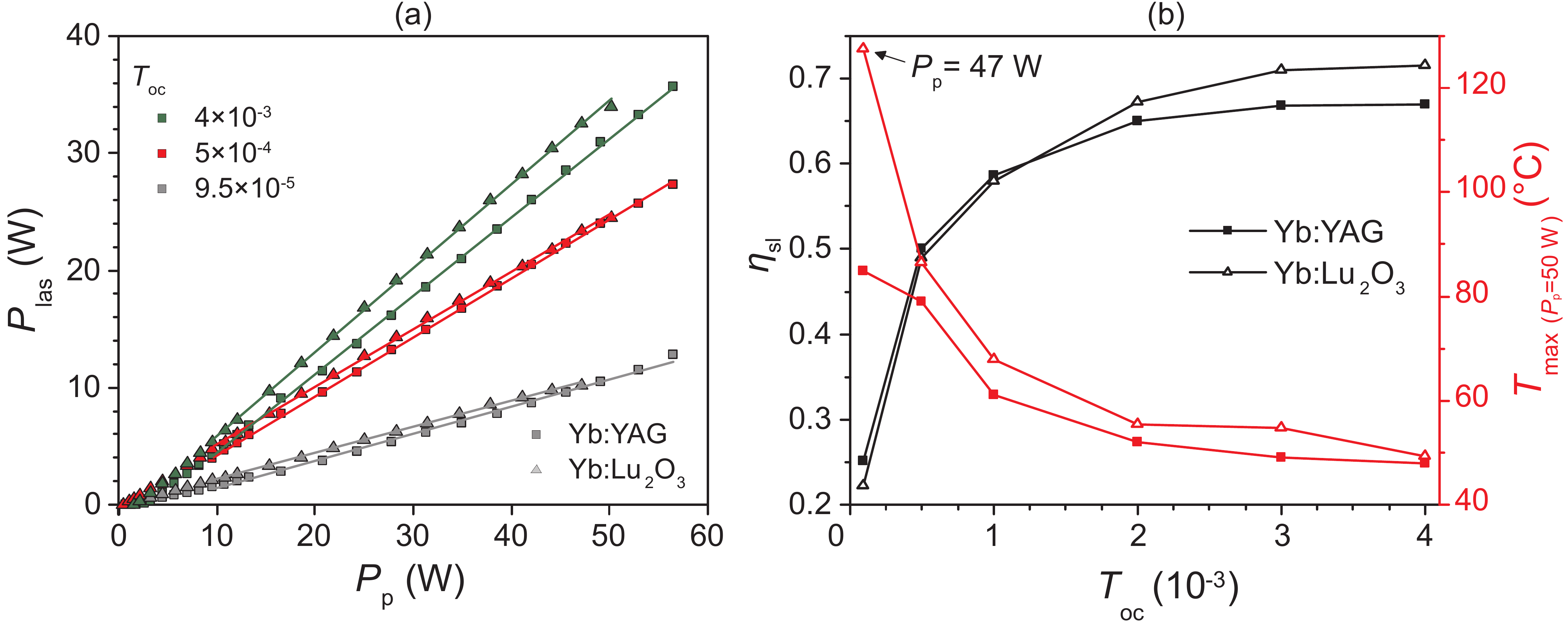} \caption{(a) Laser
      performance of a 0.22~mm thick \YAG disk and a 0.25~mm thick \LuO disk for different output
      coupler transmissions. (b) Slope efficiency and surface temperature ($P_\text{p}$ = 50~W) for
      different output coupler transmissions.}
   \label{fig:Laserperformance}
\end{figure}
In Fig.~\ref{fig:Laserperformance}\,(a) the laser performance of the \YAG and the \LuO disks
are shown for three output coupler transmissions of $9.5\times10^{-5}$, $5\times10^{-4}$, $4\times10^{-3}$.
The observed slope efficiencies as well as the measured surface temperatures for these and other
mirrors for a pump power of 50~W are shown in Fig.~\ref{fig:Laserperformance}\,(b). As the surface
temperature of \LuO exceeded a damage-critical temperature of \celsius{130} for the lowest output
coupler transmission, the maximum pump power was limited to 47~W for this mirror. For \YAG and \LuO
maximum slope efficiencies of 67~\% and 72~\% were measured at the highest output coupler
transmission rate of $4\times10^{-3}$ with a maximum output power of 36~W and 34~W, respectively. In
this configuration, the maximum optical-to-optical efficiency was 66~\% and 68~\%, respectively. It
should be noted that these ${T_\text{oc}}$ are significantly below the optimum ${T_\text{oc}}$ for
maximum laser output for these materials~\cite{Peters:OE15:7075}. Thus, the efficiencies reported
here are lower than previously reported~\cite{Kraenkel:IEEEJSelTopQuantElec}. However, from the
measured efficiencies it can be concluded that the losses through the output coupling mirror are
significantly higher than the losses in the disks at these output coupler transmissions. For both
materials, this results in low surface temperatures even at pump powers of 50~W, which range from
around \celsius{50} for large output coupler transmissions to the highest surface temperatures of
\celsius{85} and \celsius{127} for \YAG and \LuO, respectively, measured at the lowest output
coupler transmission rate of $9.5\times10^{-5}$. Here, the losses in the disks mainly cause a strong
heating of the laser disks as they dominate over the output coupler transmission losses. Despite
this relatively strong heating, for both lasers no thermal roll over could be observed, and maximum
output powers $P_\text{out}$ of 13~W and 10~W were reached, respectively. From these numbers the
intracavity power $P_\text{int}$ was derived from the known output coupler transmission
$T_\text{out}$ as
\begin{equation}
   P_\text{int}=\frac{P_\text{out}}{T_\text{oc}}.
   \label{eq:Pint}
\end{equation}
Both disks achieved their highest CW intracavity power of $137$~kW for \YAG and 105~kW for \LuO at
minimized total resonator losses, \ie, at the lowest output coupler transmission rate of
$9.5\times10^{-5}$. At such low output coupler transmissions the laser spectrum of \YAG covers
wavelengths between 1050~nm and 1085~nm, while the \LuO laser oscillates at wavelengths between
1078~nm and 1082~nm. From the laser performance at different/lowest output coupler transmissions, we derived
the resonator round-trip losses by the Caird analysis \cite{Caird:QE224:1077}. The slope efficiency
$\eta_\text{sl}$ of a laser follows from
\begin{equation}
   \eta_\text{sl} = \eta_\text{oc}\cdot\eta_\text{St}\cdot\eta_\text{abs}\cdot\eta_\text{ov}.
  \label{eq:SlopeEfficiency}
\end{equation}
with the the output coupling efficiency
$\eta_\text{oc}=\text{T}_\text{oc}/(\text{T}_\text{oc}+\text{L}_\text{int})$, the Stokes efficiency
$\eta_\text{St}$, the pump absorption efficiency $\eta_\text{abs}$, and the pump-laser mode overlap
efficiency $\eta_\text{ov}$. The product of the latter three can be abbreviated as total efficiency
$\eta_\text{tot}=\eta_\text{St}\cdot\eta_\text{abs}\cdot\eta_\text{ov}$. Trivial rearrangements lead
to
\begin{equation}
   \frac{1}{\eta_{\text{sl}}} = \frac{L_\text{int}}{\eta_\text{tot}}\cdot\frac{1}{T_\text{oc}}
   + \frac{1}{\eta_\text{tot}}
   \label{eq:Caird}
\end{equation}
which allows to derive the resonator losses from a linear fit of the inverted slope efficiencies
versus the inverted output coupler transmission. It is known that the Caird plot may not lead to
reasonable results at higher output coupler transmissions, where the slope efficiency decreases due
to loss processes at high inversion densities which are not covered by the underlying rate equations
\cite{Wolters:PHD2014}.
However, at the low output coupler transmissions used in our experiments, these effects are
negligible.
\begin{figure}
   \centering
   \includegraphics[width=\textwidth]{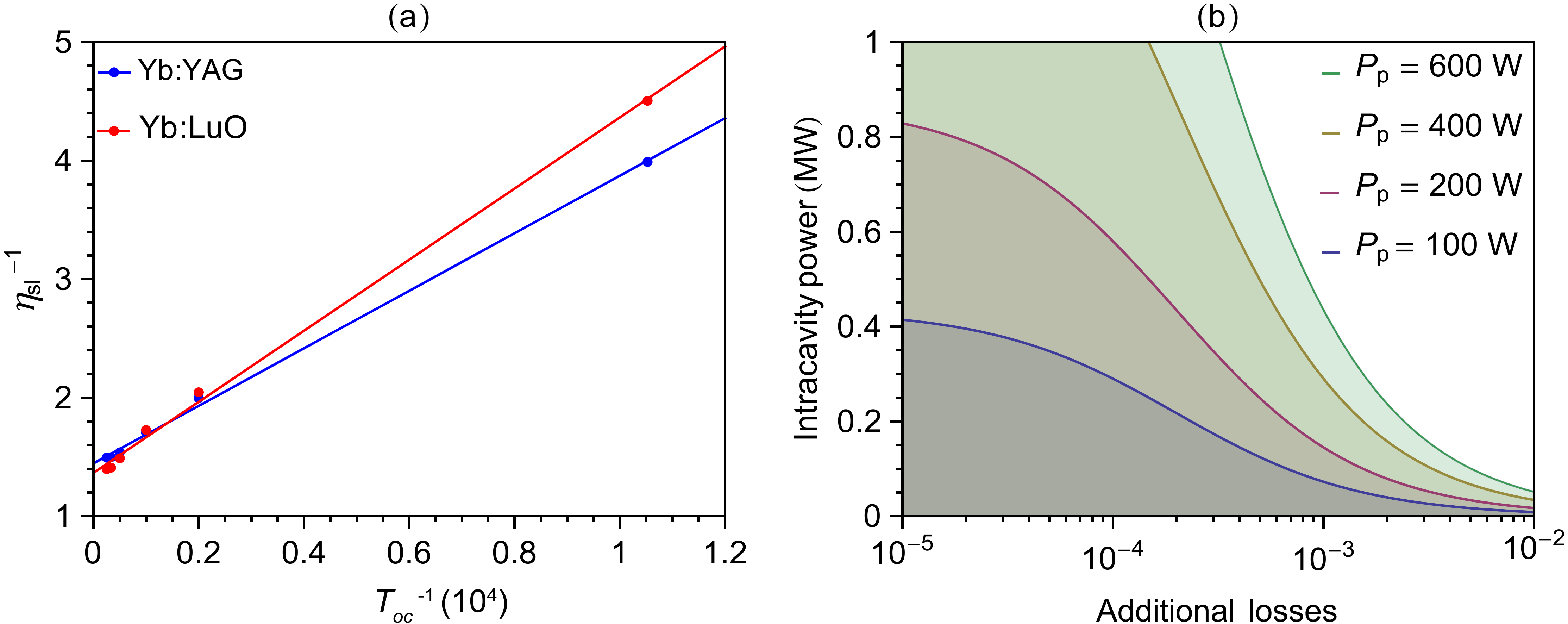}
   \caption{(a) Caird analysis of the \YAG and \LuO disk according to Eq.~\ref{eq:Caird}.
   (b) Calculated intracavity power for internal losses of $2\times10^{-4}$ for fixed pump powers against additional losses.}
   \label{fig:Calculations}
\end{figure}
The fit according to Eq.~\ref{eq:Caird} is shown in Fig.~\ref{fig:Calculations}\,(a) and results in
very low round-trip losses of $1.7\times10^{-4}$ for \YAG and only slightly higher losses of
$2.2\times10^{-4}$ for \LuO. The total efficiency amounts to 71~\% for \YAG and 77~\% for \LuO.

Considering the slope efficiencies $\eta_{\text{sl}}$ approaching the Stokes efficiencies
$\eta_{\text{st}}=\lambda_{\text{las}}/\lambda_{\text{p}}$ previously observed in
TDLs~\cite{Kraenkel:ETML11} one might argue that the absorption efficiency $\eta_{abs}$ and the
overlap efficiency of pump and laser mode $\eta_\text{ov}$ should be approaching unity. This
condition is fulfilled by deriving the intracavity losses from the ratio of the Stokes efficiency
$\eta_\text{st}$ and the slope efficiency $\eta_\text{sl}$ at the lowest output coupler transmission
from
\begin{equation}
   L_{\text{int,max}}=\left(\frac{\eta_{\text{St}}}{\eta_{\text{sl}}}-1\right)T_{\text{oc}}
   \label{eq:MaximumLosses}
\end{equation}
Despite the higher Stokes efficiency of $\eta_\text{St} \approx 0.9 > \eta_\text{tot}$ , even in
this case the corresponding losses are only slightly higher and amount to $2.4\times10^{-4}$ for
\YAG and $2.9\times10^{-4}$ for \LuO. Nota bene, this is an upper level for the intracavity losses
due to energy conservation.

The intracavity power considering realistic resonator losses of $L_{\text{int}} = 2\times10^{-4}$
plus additional losses $L_{\text{add}}$ between $10^{-5}$ and $10^{-2}$ due to output coupler
transmission or other intracavity elements required e.g. for the polarization selection were
calculated for pump powers $P_{\text{p}}$ between 100~W and 600~W with
\begin{equation}
   P_{\text{int}} = {\left(\frac{1}{L_{\text{add}}+L_{\text{int}}}\right)\cdot\eta_{\text{st}}\cdot
      (P_{\text{p}}-P_{\text{thr}})}.
   \label{eq:simplified_intracavitypower}
\end{equation}
This equation can be simplified by assuming a threshold power $P_{\text{thr}}$ of zero, which is
appropriate as pump thresholds $<0.5$~W were observed in all experiments. The resulting intracavity
powers are depicted in Fig.~\ref{fig:Calculations}\,(b).

\section{Discussion}
\label{sec:discussion}
All measurements were performed in a short linear laser resonator to allow for both, efficient
multi-mode lasing and for easy evaluation of losses, which require knowledge of the specific loss of
all optical components. Under these conditions the gain-medium-specific resonator losses
$L_{\text{int}}\le2\times10^{-4}$ can be nearly exclusively attributed to losses in the TDL
assembly, \ie, the laser-medium disk with its dielectric coatings and the metallic contacting layer.
While the \YAG disk showed in general better performance, our experiments do not provide conclusive
evidence about the material-specific advantage of \YAG or \LuO. Using a pump power of 54~W we
achieved an intracavity power of $135~\textup{kW}$ for \YAG. This is to the best of our knowledge
the highest documented CW intracavity power for a pump power lower than 100~W. Typically, such CW
intracavity powers are only achieved using pump powers of 10~kW.

The slope efficiencies obtained with \YAG and \LuO of 0.66 and 0.72, respectively are in good
agreement with previous results at such low output coupling transmissions
~\cite{Peters:OE15:7075,Fredrich:thesis:2010}. The measured laser performance at various low output
coupler transmissions below $4\times10^{-3}$ allows to precisely determine an upper limit of the
internal resonator losses of a few $10^{-4}$, which is about an order of magnitude lower than
previously assumed \cite{Fredrich:thesis:2010,Larionov:thesis:2008}. This also demonstrates the low
losses of the thin-disk laser resonator and the excellent quality of the utilized disks, which
benefited from improvements in the crystal growth and in optical coating methods over the last
decade.

The results in Fig.~\ref{fig:Calculations}\,(b) reveal that intracavity power levels in excess of
0.5~MW can be achieved at pump powers of a few 100~W, even at total resonator losses in the order of
$10^{-3}$, which would be five times higher than the intracavity losses of $2\times10^{-4}$ we
determined for multi-mode linear TDLs using state-of-the-art processed gain materials and standard
resonator mirrors. At such intracavity power levels, the required focal intensities for molecular
alignment in the order of $10^{10}$--$10^{11}~\text{W/cm}^2$ could be achieved at realistic
intracavity focal diameters of 20--40~\um. Such diameters can be obtained between two concave cavity
mirrors and do not impose a particular challenge for the resonator design. Slightly larger
intracavity foci have already been demonstrated e.g. in enhancement cavities and conventional
resonators~\cite{Carstens:OL39:2595, Zhang:OL40:1627}.

Figure 3 (b) also shows that minimized resonator losses are of crucial importance for achieving high
intracavity power levels at moderate pump power levels. We note that we operated our TDL at output
coupler transmissions in the order of $10^{-4}$. However, in the upcoming experiments the resonator
losses will be increased by additional resonator elements, for instance a Brewster plate. At
intracavity powers in excess of 100~kW even a very low transmission of 10$^{-7}$ results in leakage
of more than 10~mW, which allows for a reliable determination of the intracavity power. Moreover, we
expect the main additional losses to occur due to depolarization at polarization control elements,
\ie, reflections at Brewster elements or transmission losses at mirrors with a polarization
dependent reflectivity. Both should increase resonator extraction efficiency and thus avoid
hot-spots in the cavity. Therefore, the disk temperatures should remain lower than demonstrated
here, even when using HR mirrors.

The considerations so far were independent of the pump spot diameter on the disk. We have shown that
further scaling of the intracavity power may lead to strong heating of the disk.
This can be circumvented by choosing larger pump spot diameters, which allows to use significantly
higher pump powers. Even though the alignment sensitivity increases strongly with the laser mode
diameter on the disks~\cite{Magni:JOSAA4:1962}, TDL with centimeter-scale pump spot diameters are
reported~\cite{Beil:OE18:20712}. Even in fundamental mode operation, a 4.7~mm pump spot diameter has
been reported at a pump power of up to 830~W~\cite{Baer:OE20:7054}; this corresponds to pump power
intensities on the order of $5~\text{kW/cm}^2$. In this case the output coupler transmission was
optimized for high extraction efficiency, but at the maximum output power of 430~W the remaining
pump power deposited in the disk was still as high as 400~W. This pump power level should be
sufficient for our experiments.

The application of the laser for molecular alignment requires operation in vacuum and polarization
control. Preliminary results in vacuum ($p=5\times10^{-4}$~mbar) point towards an increased
operation stability due to the lack of atmospheric turbulence, in agreement with previous
reports~\cite{Saraceno:OE20:23535, Weichelt:SPIE2010:77210M}. Further experiments are required to
explore further challenges, \eg, the corresponding lack of convection cooling of the optical
elements and mounts. For the alignment experiments it would be sufficient to keep only the focal area
of the resonator in a vacuum chamber, but the required additional Brewster windows and polarization
optics would add complexity and increase the total resonator losses.

When operated in ultra-high-vacuum ($10^{-9}$~mbar) and combined with a continuous cold supersonic
molecular beam~\cite{Scoles:MolBeam:1}, the demonstrated laser system will allow to strongly align
and orient molecules at ``arbitrary'' repetition rates. The continuous presence of aligned molecules
will enable the envisioned application of strongly controlled molecules in modern imaging
experiments at high-repetition rate X-ray facilities~\cite{Kuepper:PRL112:083002,
Boll:PRA88:061402}.

\section{Conclusions}
\label{sec:conclusions}
We have demonstrated a thin-disk laser providing 135~kW of CW intracavity laser power. This
corresponded to an enhancement by a factor of 2500 with respect to the incident pump power of 54~W,
enabled by the low losses of state-of-the-art-processed gain disks. The internal round-trip losses
of \YAG and \LuO disks were determined to be about $2\times10^{-4}$. Calculations show that our
approach is scalable and will allow for megawatt-level CW intracavity powers, thus, enabling
field-strengths in excess of $10^{10}~\text{W/cm}^2$ in a few-10-\um-diameter focus. Such ac
electric field strengths allow for adiabatic laser alignment or mixed-field orientation of
molecules. While pulsed lasers at comparably low repetition rates have been used for this
purpose~\cite{Stapelfeldt:RMP75:543, Holmegaard:PRL102:023001, Trippel:MP111:1738,
   Kierspel:JPB:48:204002}, our approach allows for adiabatic alignment at arbitrary repetition
rates.

Coupled to a continuous molecular beam, such a setup could allow for the implementation of reactive,
chemical scattering investigations of aligned or oriented molecules~\cite{Loesch:ARPC46:555}, albeit
limited to the very small focal volume ($10^{-9}~\text{cm}^{3}$). Moreover, we point out that our
laser would also allow for the trapping of atoms and molecules using the polarizability
interaction~\cite{Friedrich:PRL74:4623, Grimm:AAMOP42:95}. It would allow for few-Kelvin deep traps
for typical small molecules~\cite{Trippel:PRA89:051401R} and even deeper traps for larger, more
polarizable molecules. This could also be utilized for \emph{in-vacuo} trapping and guiding of
nanoparticles~\cite{Ashkin:PRL24:156, Eckerskorn:OE21:30492}. The setup could be useful for many
more strong-field experiments in atomic and molecular physics, for instance, at high-repetition-rate
FELs~\cite{Mazza:NatComm5:1}.

\section*{Acknowledgments}
This work has been supported by the excellence cluster ``The Hamburg Center for Ultrafast Imaging --
Structure, Dynamics and Control of Matter at the Atomic Scale'' (CUI) of the Deutsche
Forschungsgemeinschaft (DFG EXC1074).

\begin{thebibliography}{10}
\bibitem{definition-alignment+orientation}
We distinguish between alignment and orientation according to the usual
  convention, i.\,e., alignment refers to fixing one or more molecular axes in
  space, while orientation refers to the breaking of the ``up-down'' symmetry
  along one or more axes.
\end{thebibliography}
\end{document}